# "Computer Science for all": Concepts to engage teenagers and non-CS students in technology

Bernadette Spieler[1], Maria Grandl[2], Martin Ebner[3], Wolfgang Slany[1]
[1]Graz University of Technology, Institute of Software Technology, Graz, Austria
[2]Graz University of Technology, Institute of Interactive Systems and Data Science, Graz, Austria
[3]Graz University of Technology, Department Educational Technology, Graz, Austria
bernadette.spieler@ist.tugraz.at
maria.grandl@tugraz.at
martin.ebner@tugraz.at
wolfgang.slany@tugraz.at

**Abstract:** Knowledge in Computer Science (CS) is essential, and companies have increased their demands for CS professionals. Despite this, many jobs remain unfilled. Furthermore, employees with computational thinking (CT) skills are required, even if they are not actual technicians. Moreover, the gender disparity in technology related fields is a serious problem. Even if companies want to hire women in technology, the number of women who enter these fields is remarkably low. In high schools, most teenagers acquire only low-level skills in CS. Thus, they may never understand the fundamental concepts of CS, have unrealistic expectations or preconceptions, and are influenced by stereotype-based expectations. Consequently, many teenagers exclude computing as a career path. In this research study, we present two promising concepts to overcome these challenges. First, we consider alternative paths to enter the field of CS. In 2018, a voluntary lecture "Design your own app" at the University of Graz for students of all degree programs was introduced. In total, 202 students participated. We applied a Game Development-Based Learning (GDBL) approach with the visual coding tool Pocket Code, a mobile app developed at Graz University of Technology. The students were supposed to create simple games directly on smartphones. The course received positive evaluations and led to our second concept; In January 2019, we started to design a MOOC (Massive Open Online Course) with the title "Get FIT in Computer Science". The MOOC will be launched in August 2019 on the platform iMooX.at and will provide a general introduction to the field of CS. For exercises and the final submission, the students need to apply game design strategies by using Pocket Code. The MOOC has several target groups. First, this course can be used to encourage young women who have little to no previous knowledge in CS. Second, it should help all teenagers to get a more realistic picture of CS to its basic concepts. Third, teachers can use the course materials to lead high school classes (Open Educational Resources). Finally, the MOOC can be accessed by everyone interested in this topic, thus students of other majors can acquire CS skills.

## 1. Introduction

IT specialists are needed worldwide in order for businesses to be competitive in science and technology. Thus, there is a growing demand for IT professionals (Cuff, 2015). On the one hand, the enrolment of students in Computer Science (CS) programs according to Informatics Europe (2015) in European countries in general has slightly increased. On the other hand, degree courses in computer science have very high drop-out rates (European Commission, 2015). There are many reasons why students drop out of university, but this report pointing out that students who have no previous knowledge in CS are more likely to drop out in comparison to students who already have programming experience and a basic understanding of the important concepts of CS. In addition, data from European Statistics Eurostat (2019) confirm a low percentage of female students in studies related to Information and Communication Technologies (ICT). In Austria, the percentage is only about 14%, which is even lower than the EU average (17%), but higher than the share of women in ICT studies in Luxembourg (10%), Belgium (8%) and in the Netherlands (6%). Female students who decide to enter the field of ICT or CS are then confronted with various challenges and prejustices. During the study of CS, an already homogeneous group of male students tends to become more and more similar, or rather, those who have a different mindset tend to drop out. In STEM (Science, Technology, Engineering and Math's) subjects in Austria, all male and female students in bachelor programs have the same success rates of about one third (Binder et al., 2017). However, more men (16%) further enrol for a graduate study (master's or doctoral program) than women (10%). Degree programs of other, non-STEM fields are more likely to be completed by women. Here, it is apparent that the higher the proportion of women in a subject, the more frequently women complete their studies compared to men (Dormayr and Winkler, 2016). The only exception in success rates and gender is in the field of computer science: Here, women have a 10%-points lower success rate than men. One explanation of this gender related success rate is the differences in the school education. Overall, women in STEM studies more often have a degree from a general academic secondary school (women 66% vs. men 50%) and more rarely completed their secondary education at a technical colleges than men (women 8% vs. men 37%). To conclude, the success rate of female students in the STEM area with technical college's degree is much higher than those of students with a general academic secondary school degree.

The percentage of female students at Graz University of Technology (TU Graz) increases slightly every year but it is mainly due to the high amount of female first semester students in architecture, biomedical engineering, and chemistry. Details of the percentage of women's bachelor's degrees 2017/2018 are presented in Figure 1.

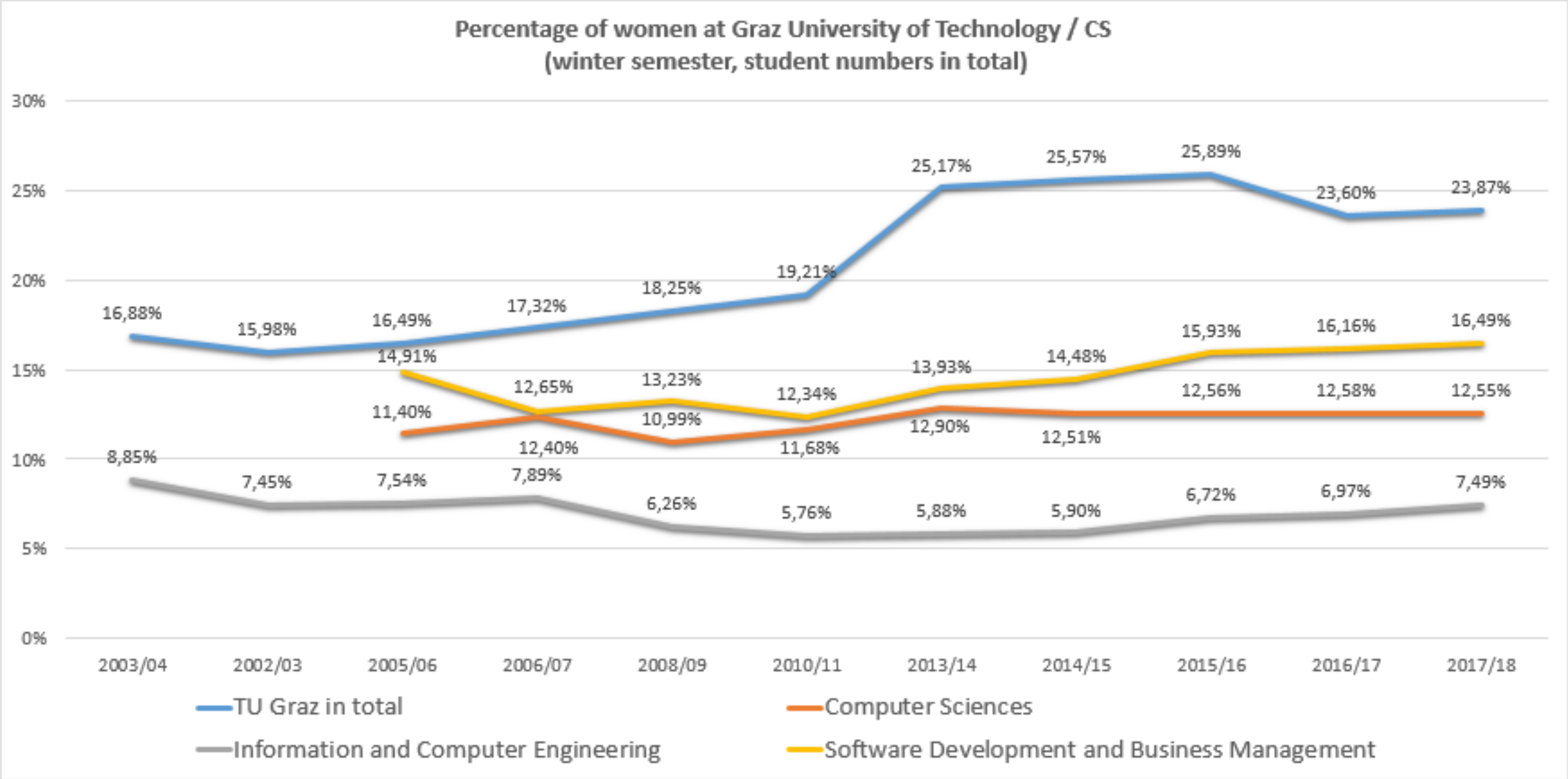

**Figure 1:** Percentage if of Computer Science degree programs at TU Graz (winter term of 2003/04 to 2017/18). [online] https://online.tugraz.at/tugonline/Studierendenstatistik.html

To conclude, the situation is not surprising as it is similar to those in other developed countries. Reasons why female teenagers decide not to choose CS as a major are diverse, but there are general issues that must be addressed so that alternative approaches can be implemented in order to increase female achievement (Cheryan et al., 2013). In addition to this incorrect assumption, there is a major lack of exposure to CS at schools all over Europe (CECE, 2015).

This paper is organized as follows: First, the paper provides a literature review by presenting important eLearning and MOOCs concepts. The advantages of game development challenges and the importance of more diverse people in CS are described in more detail. Subsequently, Section 3 presents results from our lecture, "Design your own App", an elective lecture for all students at the public university Uni Graz. In Section 4, we show the development of the MOOC (Massive Open Online Courses) with the title "Get FIT in Computer Science" which is based on the lecture. Both concepts include game development challenges as a key component and are conceptualized and developed in parallel. Sections 5 and 6 conclude this paper and describe the author's future work.

## 2. Related Literature

A recent New York times article by Singer (2019) stated that a growing amount of students with no computing experiences feel motivated to learn about coding but are not willing to study CS as a major. On the one hand, learning computing skills can be a fast path to employment and jobs in that area are well-paid and the demand is unbounded. On the other hand, knowledge in computing is essential and computational and problem solving skills are required even if you are not working as a programmer. If universities treat computer science more as a course that lays the foundations for other courses such as physics, students could learn important computer skills and then apply them to other courses while keeping their career prospects open. In this section, related literature to the issues emerged are presented.

## 2.1 Technology Enhanced Learning and Massive Open Online Courses (MOOCs)

For students who want to learn about ICT or CS, eLearning concepts, online courses, or MOOCs, in particular, are a perfect way to start and to support distance education or lifelong learning. The rise of MOOCs can be dated back to the so-called "year of the MOOCs" in 2012 (Pappana, 2012). Since then, more and more higher education institutions began to produce so called xMOOCs (Carson and Schmid, 2012), by using large MOOC platforms like Udacity, edX, or Coursera to publish their video-based courses or lectures. It has become apparent that those MOOCs helped to reach a broad public and to introduce new didactic approaches (Ebner and Schön, 2019). However, the most well known problems of MOOCs are the high drop-out rates (Khalil and Ebner, 2014) and the difficulty in certifying courses in the context of a higher education institution (Kopp and Ebner, 2017).

In 2014, TU Graz and the Uni Graz founded the first and currently only MOOC-platform in Austria, called iMooX (Khalil and Ebner, 2016). Following the idea of open education, each single learning object within a course on iMooX is published with an open license and can be identified as an Open Educational Resource (OER). Due to a very strict copyright law in Austria, a reuse of the course materials for educational purposes would be difficult (Ebner et al., 2016). Therefore, it is of high importance to produce and offer all the content as OER. In recent years, more than 60 MOOCs have been offered on iMooX.at, addressing people from different educational sectors with different interests.

In the context of this research work, one of the most relevant courses was the so-called "Mathe-FIT-MOOC" (Get-FIT-In-Math-MOOC). This MOOC aims to act as a preparatory course to bridge the knowledge gap in Math between school and university level. Following the concept of flipped classroom, the students had to finish this 6-week course before the start of the first semester. They were supposed to study the online content (videos and exercises) in August and September. After that, it was possible to attend the corresponding lecture, where exercises were discussed in lecture halls at the university The final examination, a multiple-choice questionnaire, was done right before the start of the first semester. The idea of this course is to better prepare beginners of science-oriented studies for the challenges of their first semester at university.

## 2.2 Digital Literacy versus Computer Science

A critical approach to new technologies requires a general understanding of the logical and technical aspects behind them. However, European school systems are mostly concerned with Digital Literacy and Information and Communication Technology (ICT) as a supporting context-free medium/technology to enhance learning." (CECE, 2017). On the one hand, pupils should learn how to operate a computer in an efficient way, with a focus on standard software tools and services of the World Wide Web. On the other hand, CS is more than just using the computer as a tool. It is a "distinct scientific discipline" that plays an insignificant and insufficient role in the school curricula of most European school systems (ibid.). Consequently, CS remains a great unknown for many pupils worldwide. The "Hour of Code", a global event organized by the non-profit organization Code.org, is one example of many initiatives with the goal, to help people, especially children and teenager, to understand the fundamental concepts of CS.

Computational Thinking (CT) is considered an essential skill in the 21st century. According to J. Wing, it "represents a universally applicable attitude and skill set everyone, not just computer scientists, would be eager to learn and use." (Wing, 2006). Under certain conditions, coding activities are one way to train CT, by which we understand the "thought processes involved in formulating problems and their solutions so that the solutions are represented in a form that can be effectively carried out by an information-processing agent" (Wing, 2010). Many countries around the world, including Austria and Germany, organize an annual Bebras contest, that usually takes place in November (Brebas Challenge Austria: http://wettbewerb.biber.ocg.at/, Germany: https://bwinf.de/biber/, UK: http://www.bebras.uk/). The aim of this contest is to help pupils of different age groups to develop computational thinking skills, to boost interest in computer science, to think about tricky tasks related to computer science and to disseminate basic concepts of computer science without the actual use of a computer.

## 2.3 Game Development Based Learning (GDBL) Challenges & Tools

Game Development Based Learning (GDBL) challenges are popular in introductory programming courses at all educational levels (Wu and Wang, 2012). These educational contexts provide engaging, goal-oriented, and interactive tasks for lectures, thereby supporting the transfer of knowledge in a fun and pedagogic manner (Romero, 2012; Khaleel et. al., 2015). Thus, the lecture becomes more dynamic, collaborative, and attractive for students. This form of learning is a popular strategy to motivate and engage students. The active involvement of students as designers and producers of mobile games can have an even greater potential for empowering students by encouraging critical thinking and improving learning outcomes as well as creativity, problem-solving, and critical thinking skills (Ya-Ting et al., 2013; Vos, van der Meijden and Denessen, 2011). Actions to promote CS should improve the quality and relevance of CS education to enable (young) people to make better career choices, get quality jobs and improve their lives in the end (European Commission, 2016).

New technologies and tools influenced the ways of learning and teaching in the 21st century (Kahn, 2017). Game Development-based approaches could be easily applied with block-based visual oriented programming tools, like Scratch (https://scratch.mit.edu/), Snap (https://snap.berkeley.edu/), or AppInventor (http://appinventor.mit.edu). These tools have all had very similar goals: they focus on younger learners, support novices in their first programming steps, and provide a block-based visual-oriented programming language which allows students to recognize blocks instead of recalling syntax (Tumlin, 2017). In 2010, the Catrobat team (https://catrobat.org) at TU Graz started to develop an educational app with the purpose of showing teenagers an interesting and engaging tool for learning programming. One of our apps is Pocket Code (free on Google Play: https://catrob.at/pc and iTunes: https://catrob.at/iosPC), a visual programming language environment that allows the creation of games, stories, animations, and many types of other apps directly on smartphones or tablets. With the use of simple graphic blocks, users can create apps directly on the mobile phone, see Figure 2.

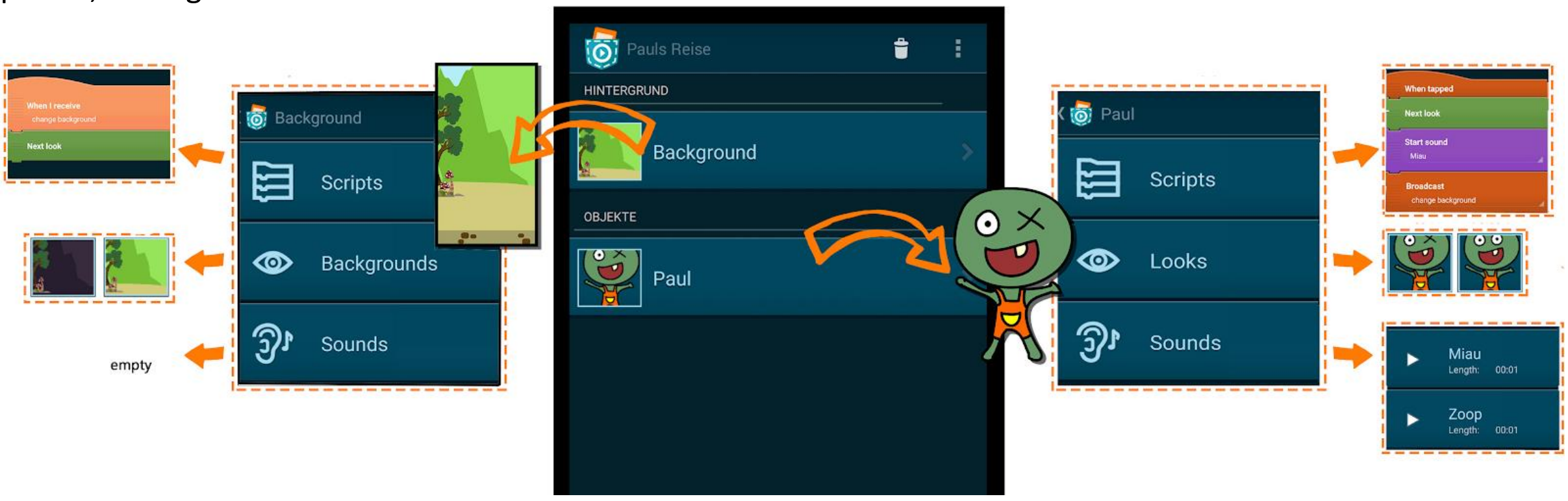

**Figure 2:** Pocket Code's UI.

## 2.4 Diversity in Computer Science

The introduction section stated the problem: The percentages of female university students in ICT fields in Austria currently varies between 4% to 20%, with percentages dropping faster for female students over the number of their study years compared to male students (Binder et al., 2017). The demand for skilled computer science professionals is growing, but the faltering rate of female students and low amount of female professionals currently in CS illustrates a troubling disparity of gender in an otherwise promising field.

Many researchers conclude that we do not need to "fix the women", we need to fix the system (e.g., Alvardo, Cao, and Minnes, 2017; Master, Sapna, and Meltzoff, 2016). Predominant stereotypes, and missing role models (Young et al., 2013), as well as other social and cultural factors (e.g., gender role socialization, peer groups), expose technology as male-dominated field. Girls' lower sense of belonging corresponds with the feeling of "Lack of Fit" or "the Sense of Not Fitting In" with computer science stereotypes. This occurs if female CS students feel that they do not receive help, question their ability in CS, or feel intimidated by others. If the profession does not fit to the "traditional gender model", one is not as likely to pursue or feels discriminated against by someone who does. To be socially connected and respected is a strong initial motivator (Baumeister and Leary, 1995; Walton and Cohen, 2007): it can "create a sense of belonging that can reinforce students' self-efficacy and connections to community that support student perceptions of their ability within the field" (Veilleux et al., 2013, p. 64). This is important in students' decision to pursue IT and helps to identify with the field (Beyer, 2016). Since the stereotype in IT is more associated with a male role, female teenagers are less likely to feel a sense of belonging with these stereotypes (Master, Sapna, and Meltzoff, 2016; Cheryan, 2012). All of these factors make young women question whether their abilities and interests are harmonious with their selected field.

## 3. Results that informed the MOOC Course

In the summer term of 2018, we introduced for the first time the obligatory lecture “Design your own App” for all bachelor's, master’s, and doctoral students at the University of Graz. The lecture was a cooperation between TU Graz and University Graz and was promoted in different ways, either through the cooperation program by University Graz (mailing list, Facebook, website) or by TU Graz (posters and flyers). A total of 202 students from diverse degree programs attended the first lecture. In the first lecture, before providing any additional information, an online survey was conducted anonymously. A total of 135 students filled out the survey, see Figure 3.

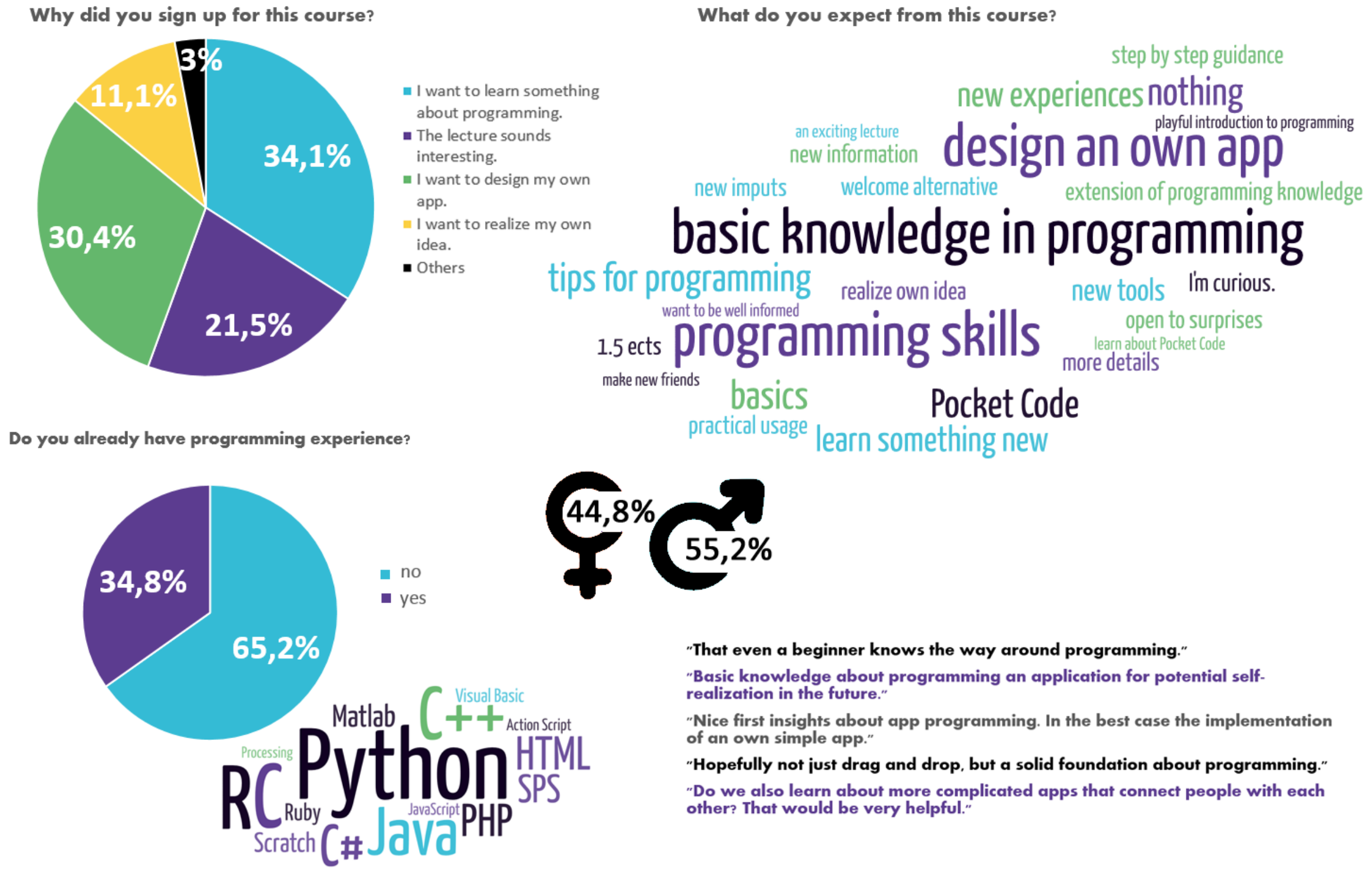

**Figure 3:** Expectations of the course “Design your own app 2018”.

In addition, we knew the majors of the participants. The majority of students, a total of 43.82%, came from three degree programs: 17.98% were students from teacher training studies (e.g., for languages, math’s, geography, physics, etc.), 15.73% were students from Business Management and Economics, or 10.11% beginner students from Computer Science (these were first semester students from TU Graz and programming beginners). A percentage of 56.18% came from 120 different degree programs, including psychology, legal sciences, molecular biology, sociology, biology, geography, history, chemistry, or philosophy. This variety of students was impressive. This correlates with the literature review (see section 2.4) which stated that there is a huge demand of students from different majors to learn something meaningful about programming and CS.

The goal of the lecture was to create a playful and easy entry point into the world of programming and to create a personalised gaming app with the Pocket Code tool. This interactive lecture combined several concepts, e.g.,

- Stimulating computational thinking skills, such as thinking abstractly, and the deconstruction of a problem into smaller pieces, e.g., by using logical puzzles
- Moving beyond participation and creative thinking via game-making and coding

- Using new tools and allowing students to be creative

Students had 10 units at one hour each. All lectures followed a similar structure. At the beginning, they received a 30-minute lecture explaining theoretical input (e.g., history of CS, game design principles, etc.) followed by a 15-minute hands-on activity with Pocket Code (kind of "small challenges"). At the end, voluntary students were able to present their challenge. From unit to unit, they received a small homework example to be completed in Pocket Code. After unit four, the groups for the final projects were formed (students with different majors were assigned randomly to groups of two). For the final project, students could choose three out of four tasks: 1) create a short video (who are you? what is your project about?), 2) design your own gaming app with Pocket Code (this task was mandatory), 3) share your knowledge (teach Pocket Code to someone from our main target group: teenagers between 11-17 years old), and 4) write a report, e.g., a lesson plan to integrate Pocket Code in school classes or a "lesson learnt" review. For task 2 - the app - the student had to fulfill the following requirements (Spieler and Slany, 2018):

- choose a genre, theme and goal of the game
- integrate Shape of a Game
- use different MDAs (min. 2)
- choose two diversifiers, e.g., two sensors, integrate a quiz

For this final project, 20 hours per student was allotted. The submission was done through our Catrobat community page.

The lecture was very successful in showing a diversity of students from different majors a playful and gamified approach to learn programming. Thus, we planned to continue with the lecture in the summer term of 2019. This time, we want to evaluate students gained computational thinking skills as well. This will be done via a pre-test and post-test, which consists of different questions from the Brebas challenge (see Section 2.3). Furthermore, as stated in the introduction, at TU Graz the numbers of female students in computer science are alarmingly low and the literature supports the argument to provide them with prior opportunities to learn CS literacy. Thus, based on the lecture done in 2018, in January 2019 we started a concept for a MOOC course for Computer Science.

## 4. Development of a MOOC Course to "Get FIT in Computer Science"

In January 2019, we started to design a MOOC with the title "Get FIT in Computer Science", which is based on the experiences and feedback gained from the obligatory lecture "Design your own App" and follows the ideas of the MOOC created for Math. It will provide a general introduction to the field of CS and programming. Besides videos, the MOOC will include interactive exercises and programming tasks, where beginners are supposed to apply game design strategies by using Pocket Code. The MOOC will be launched on August 5th, 2019 on iMooX.at and will consist of seven modules, which are presented in Figure 4. Course registration is already open: https://imoox.at/mooc/go/Info-Fit19

Every module of the MOOC will include an exercise (e.g. write pseudocode, execute assembler commands, convert decimal into binary numbers and vice versa) and self-assessment questions in a multiple-choice format. To attract more female students, we tried to counteract stereotypes and clarify expectations. First of all, all the explanations in the MOOC are done by women. Secondly, we created a video to talk about famous female (computer) scientists in the first module. As well as that, we decided to use a game design approach to point out that programming is a highly creative process.

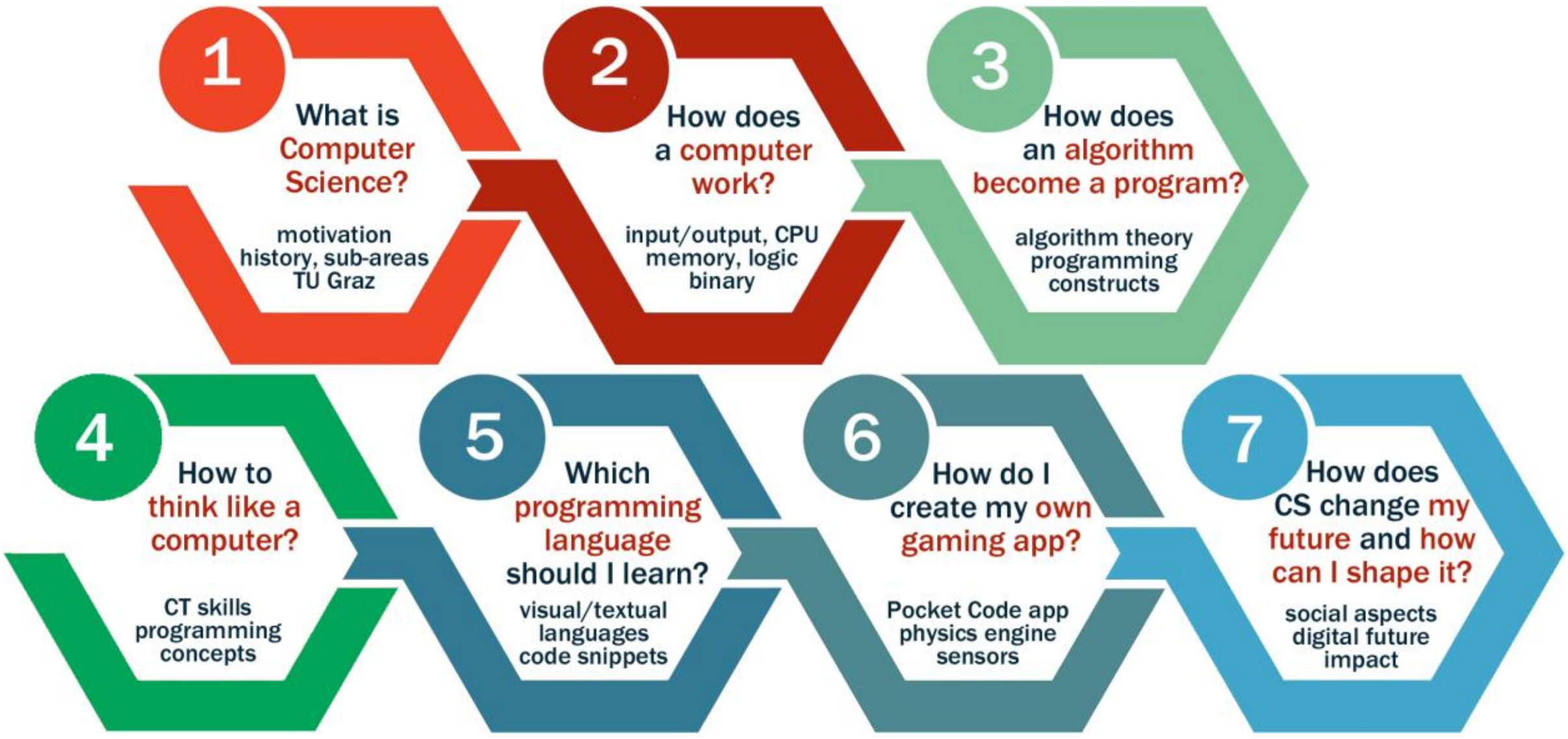

**Figure 4:** The MOOC "Get FIT in Computer Science" consists of seven modules. The content of each module is represented by a central question.

After the completion of the MOOC, students will have the possibility to attend an one week lecture (2 hours per day) in September 2019, with the goal to do programming lessons together, ask questions and to work on their final submission (i.e., the Pocket Code gaming app) at the end of September 2019. This concept is very similar to the previous described Get-Fit-In-Math-MOOC (see Section 2.1)

## 5. Conclusion & Outlook

Just working towards the goal of acquiring more women for technology will not improve the overall situation and will not automatically changes the culture in information technologies (Frieze and Quesenberry, 2015). Inclusive environments are the key; we must consider alternative routes to IT and multiple points of entry. The technology sector needs people who can design, develop, analyze, and manage information technology rather than people who are mainly developers who do objective assessment of technology or do just programming tasks (Cukier et al., 2002). Gender equality means that all humans should have the same opportunities in order to ensure a homogeneous set of experiences among the workforce (Stout and Camp, 2014). The MOOC course and the lecture are two promising concepts to attract a group of students which is more diverse. For further evaluations students have to fill out a questionnaire at the end of the MOOC course. In addition, we will ask CS students who attended the MOOC after the lecture "Programming 0" which is a first semester lecture at TU Graz if the MOOC helped them in applying concepts during their course of study.